\documentclass[preprints,article,accept,moreauthors,pdftex]{Definitions/mdpi} 
\usepackage{comment} 
\firstpage{1} 
\pubvolume{13}
\issuenum{9}
\articlenumber{1591}
\pubyear{2021}
\copyrightyear{2021}
\datereceived{23 July 2021} 
\dateaccepted{17 August 2021} 
\datepublished{29 August 2021} 
\hreflink{https://doi.org/10.3390/sym13091591} % If needed use \linebreak
\pdfoutput=1

\renewcommand{\mu}{\upmu}
\renewcommand{\gamma}{\upgamma}
\newcommand{\meg}{{\mu}^+ \rightarrow {\rm e}^+ \gamma}
\newcommand{\megc}{\ifmmode\mathrm{\mu^+ \to e^+ \gamma}\else$\mathrm{\mu^+ \to e^+ \gamma}$\fi}
\newcommand{\mutoe}{{\mu}^- N \rightarrow {\rm e}^- N}
\newcommand{\mutoeee}{\mu \rightarrow 3{\rm e}}
\newcommand{\tautolgamma}{\uptau \rightarrow \ell \gamma}
\newcommand{\tautolll}{\uptau \rightarrow \ell \ell \ell}

\newcommand*{\egamma}         {E_{\gamma}}
\newcommand*{\ppositron}      {p_{{\rm e}^+}}
\newcommand*{\Thetaegamma}    {\Theta_{{\rm e}^+ \gamma}}
\newcommand*{\tegamma}        {t_{{\rm e}^+ \gamma}}
\newcommand{\michel}{\mu^+ \to {\rm e}^+ \upnu\bar{\upnu}}
\newcommand{\radiative}{\mu^+ \to {\rm e}^+ \upnu\bar{\upnu}\gamma}
\newcommand{\aif}{{\rm e}^+ {\rm e}^- \to \gamma \gamma}\newcommand{\convme}{\ifmmode\mathrm{\mu^- \to e^-}\else$\mathrm{\mu^- \to e^-}$\fi}
\newcommand{\mute}{\ifmmode\mathrm{\mu^+ \to {rm e^+e^+e^-}}\else $\mathrm{\mu^+ \to e^+e^+e^-}$\fi}
\Title{The Search for $\meg$ with 10$^{-14}$ Sensitivity: the Upgrade of the MEG Experiment}

\TitleCitation{The Search for $\meg$ with 10$^{-14}$ Sensitivity: the Upgrade of the MEG Experiment}

\newcommand*{\INFNPi}{INFN Sezione di Pisa$^{a}$; Dipartimento di Fisica$^{b}$ dell'Universit\`a, Largo B.~Pontecorvo~3, 56127 Pisa, Italy}
\newcommand*{\INFNGe}{INFN Sezione di Genova$^{a}$; Dipartimento di Fisica$^{b}$ dell'Universit\`a, Via Dodecaneso 33, 16146 Genoa, Italy}
\newcommand*{\INFNPv}{INFN Sezione di Pavia$^{a}$; Dipartimento di Fisica$^{b}$ dell'Universit\`a, Via Bassi 6, 27100 Pavia, Italy}
\newcommand*{\INFNRm}{INFN Sezione di Roma$^{a}$; Dipartimento di Fisica$^{b}$ dell'Universit\`a ``Sapienza'', Piazzale A.~Moro 2, 00185 Rome, Italy}
\newcommand*{\INFNLe}{INFN Sezione di Lecce$^{a}$; Dipartimento di Matematica e Fisica$^{b}$ dell'Universit\`a del Salento, Via per Arnesano, 73100 Lecce, Italy}
\newcommand*{\ICEPP} {ICEPP, The University of Tokyo, 7-3-1 Hongo, Bunkyo-ku, Tokyo 113-0033, Japan }
\newcommand*{\UCI}   {University of California, Irvine, CA 92697, USA}
\newcommand*{\KEK}   {KEK, High Energy Accelerator Research Organization 1-1 Oho, Tsukuba, Ibaraki 305-0801, Japan}
\newcommand*{\PSI}   {Paul Scherrer Institut PSI, 5232 Villigen, Switzerland}
\newcommand*{\BINP}  {Budker Institute of Nuclear Physics of Siberian Branch of Russian Academy of Sciences, 630090 Novosibirsk, Russia}
\newcommand*{\JINR}  {Joint Institute for Nuclear Research, 141980 Dubna, Russia}
\newcommand*{\ETHZ}  {Swiss Federal Institute of Technology ETH, 8093 Zurich, Switzerland}
\newcommand*{\NOVST} {Novosibirsk State Technical University, 630092 Novosibirsk, Russia}

\newcommand*{\INFNPiRef}{1}
\newcommand*{\JINRRef}{2}
\newcommand*{\INFNGeRef}{3}
\newcommand*{\INFNPvRef}{4}
\newcommand*{\INFNRmRef}{5}
\newcommand*{\INFNLeRef}{6}
\newcommand*{\BINPRef}{7}
\newcommand*{\NOVSTRef}{8}
\newcommand*{\PSIRef}{9}
\newcommand*{\ICEPPRef}{10}
\newcommand*{\UCIRef}{11}
\newcommand*{\KEKRef}{12}
\newcommand*{\ETHZRef}{13}

\foreach \x in {a, ..., z}{%
\expandafter\xdef\csname orcid\x\endcsname{\noexpand\href{https://orcid.org/\csname orcidauthor\x\endcsname}{\noexpand\orcidicon}}
}

\Author{
The MEG II Collaboration\\
       Alessandro~M.~Baldini~$^{\INFNPiRef a}$\orcidg{},
       Vladimir~Baranov~$^{\JINRRef}$,
       Michele~Biasotti~$^{\INFNGeRef ab}$,
       Gianluigi~Boca~$^{\INFNPvRef ab}$\orcids{},
       Paolo~W.~Cattaneo~$^{\INFNPvRef a}$\orcidr{},
       Gianluca~Cavoto~$^{\INFNRmRef ab}$\orcidP{},
       Fabrizio~Cei~$^{\INFNPiRef ab}$\orcidh{},
       Marco~Chiappini~$^{\INFNPiRef ab}$\orcidi{},
       Gianluigi~Chiarello~$^{\INFNRmRef ab}$\orcidQ{},
       Alessandro~Corvaglia~$^{\INFNLeRef ab}$\orcidv{},
       Federica~Cuna~$^{\INFNLeRef ab}$\orcidw{},
       Giovanni~dal~Maso~$^{\INFNPiRef ab, \PSIRef}$\orcidf{},
       Antonio~de~Bari~$^{\INFNPvRef ab}$\orcidt{},
       Matteo~De~Gerone~$^{\INFNGeRef a}$\orcidC{},
       Marco~Francesconi~$^{\INFNPiRef ab}$\orcidl{},
       Luca~Galli~$^{\INFNPiRef a}$\orcidn{},
       Giovanni~Gallucci~$^{\INFNGeRef a}$\orcidE{},
       Flavio~Gatti~$^{\INFNGeRef ab}$\orcidD{},
       Francesco~Grancagnolo~$^{\INFNLeRef a}$\orcidx{},
       Marco~Grassi~$^{\INFNPiRef a}$\orcidm{},
       Dmitry~N.~Grigoriev~$^{\BINPRef,\NOVSTRef}$,
       Malte~Hildebrandt~$^{\PSIRef}$\orcidb{},
       Kei~Ieki~$^{\ICEPPRef}$,
       Fedor~Ignatov~$^{\BINPRef}$,
       Toshiyuki~Iwamoto~$^{\ICEPPRef}$\orcidH{},
       Peter-Raymond~Kettle~$^{\PSIRef}$,
       Nikolay~Khomutov~$^{\JINRRef}$,
       Satoru~Kobayashi~$^{\ICEPPRef}$\orcidI{},
       Alexander~Kolesnikov~$^{\JINRRef}$,
       Nikolay~Kravchuk~$^{\JINRRef}$,
       Victor~Krylov~$^{\JINRRef}$,
       Nikolay~Kuchinskiy~$^{\JINRRef}$,
       William~Kyle~$^{\UCIRef}$\orcidN{},
       Terence~Libeiro~$^{\UCIRef}$,
       Vladimir~Malyshev~$^{\JINRRef}$,
       Manuel~Meucci~$^{\INFNRmRef ab}$\orcidR{},
       Satoshi~Mihara~$^{\KEKRef}$\orcidF{},
       William~Molzon~$^{\UCIRef}$\orcidM{},
       Toshinori~Mori~$^{\ICEPPRef}$\orcidJ{},
       Alexander~Mtchedlishvili~$^{\PSIRef}$,
       Mitsutaka~Nakao~$^{\ICEPPRef}$,
       Donato~Nicol\`o~$^{\INFNPiRef ab}$\orcido{},
       Hajime~Nishiguchi~$^{\KEKRef}$\orcidG{},
       Shinji~Ogawa~$^{\ICEPPRef}$,
       Rina~Onda~$^{\ICEPPRef}$\orcidK{},
       Wataru~Ootani~$^{\ICEPPRef}$,
       Atsushi~Oya~$^{\ICEPPRef}$,
       Dylan~Palo~$^{\UCIRef}$\orcidO{},
       Marco~Panareo~$^{\INFNLeRef ab}$\orcidy{},
       Angela~Papa~$^{\INFNPiRef ab, \PSIRef}$\orcide{},
       Valerio~Pettinacci~$^{\INFNRmRef a}$\orcidS{},
       Alexander~Popov~$^{\BINPRef}$,
       Francesco~Renga~$^{\INFNRmRef a}$\orcidT{},
       Stefan~Ritt~$^{\PSIRef}$\orcidc{},
       Massimo~Rossella~$^{\INFNPvRef a}$\orcidu{},
       Aleksander~Rozhdestvensky~$^{\JINRRef}$,
       Patrick~Schwendimann~$^{\PSIRef,\ETHZRef}$,
       Kohei~Shimada~$^{\ICEPPRef}$,
       Giovanni~Signorelli~$^{\INFNPiRef a}$\orcidq{},
       Alexey~Stoykov~$^{\PSIRef}$\orcida{},
       Giovanni~F.~Tassielli~$^{\INFNLeRef ab}$\orcidz{},
       Kazuki~Toyoda~$^{\ICEPPRef}$,
       Yusuke~Uchiyama~$^{\ICEPPRef,}$*\orcidA{},
       Masashi~Usami~$^{\ICEPPRef}$,
       Cecilia~Voena~$^{\INFNRmRef a,}$*\orcidB{},
       Kosuke~Yanai~$^{\ICEPPRef}$,
       Kensuke~Yamamoto~$^{\ICEPPRef}$,
       Taku~Yonemoto~$^{\ICEPPRef}$
       and	
       Yury~V.~Yudin~$^{\BINPRef}$
}

\address{\INFNPiRef \quad \INFNPi\\
         \JINRRef   \quad  \JINR\\
         \INFNGeRef \quad \INFNGe\\
         \INFNPvRef \quad \INFNPv\\
         \INFNRmRef \quad \INFNRm\\
         \INFNLeRef \quad \INFNLe\\ 
         \BINPRef   \quad \BINP\\
         \NOVSTRef  \quad \NOVST\\
         \PSIRef    \quad \PSI\\ 
         \ICEPPRef  \quad \ICEPP\\
         \UCIRef    \quad \UCI\\
         \KEKRef    \quad \KEK\\
         \ETHZRef   \quad \ETHZ
}

\AuthorNames{Alessandro M. Baldini, Vladimir Baranov, Michele~Biasotti et al.}

\AuthorCitation{Baldini, A. M.; Baranov, V.; Biasotti, M. et al.}

\corres{Correspondence: cecilia.voena@roma1.infn.it; Tel.: +39-06-49914268 (C.V.), uchiyama@icepp.s.u-tokyo.ac.jp; Tel.: +81-3-3815-8384 (Y.U.)}

\abstract{The MEG experiment took data at the Paul Scherrer Institute in the years 2009--2013 
to test the violation of the lepton flavour conservation law, which originates from an accidental symmetry that the Standard Model of elementary particle physics has,
 and published the most stringent limit on the charged lepton flavour violating decay $\meg$: BR($\meg$) $<4.2 \times 10^{-13}$ at $90\%$ confidence level.
The MEG detector has been upgraded in order to reach a sensitivity of  $6\times10^{-14}$.
The basic principle of MEG II is to achieve the highest possible sensitivity using the full muon beam intensity at the Paul Scherrer Institute ($7\times10^{7}$ muons/s) with an upgraded detector.
The main improvements are better rate capability of all sub-detectors and improved resolutions while keeping the same detector concept.
In this paper, we present the current status of the preparation, integration and commissioning of the MEG II detector in the recent engineering runs.}


\begin{document}
%%%%%%%%%%%%%%%%%%%%%%%%%%%%%%%%%%%%%%%%%%
%\setcounter{section}{-1} %% Remove this when starting to work on the template.
\section{Introduction}
\label{sec:introduction}
Lepton flavour conservation is associated with an approximate symmetry in the Standard Model (SM) of elementary particle physics owing to the tiny neutrino masses. It is not protected by an explicit gauge symmetry and could be broken in New Physics (NP) beyond the SM. Because of the approximate symmetry, the processes of charged lepton flavour violation (CLFV) are allowed only at extremely small branching ratios ($\ll 10^{-50}$) in the SM.
%Charged Lepton Flavour Violation (CLFV) processes are allowed in the Standard Model (SM) of elementary particle physics, accounting for measured neutrino mass differences and mixing angles, at extremely small branching ratios ($\ll 10^{-50}$). 
Therefore, they are free of SM background and they are ideal probes for NP searches. CLFV transitions have been searched for in a variety of channels, but no evidence has been found so far. Nonetheless, they are predicted at measurable rates, not far from present experimental limits, by many SM extensions \cite{Cali2018} and in particular in light of the recent muon $g-2$ precision measurement \cite{gminus2:2021} at Fermilab, for example in \cite{Calibbi2021} (flavour symmetries) and in \cite{Lindner2018} (several extensions of SM including MSSM).

Among the CLFV processes, the decay $\meg$ is very sensitive to NP. The MEG experiment at the Paul Scherrer Institute (PSI) searched for this decay in the period 2009--2013 and set the current best world limit to its branching ratio: $BR(\meg) < 4.2 \times 10^{-13}$ at 90$\%$ confidence level \cite{meg2016}.

Other CLFV channels currently investigated are $\mutoe$, $\mutoeee$, $\tautolgamma$ and $\tautolll$ ($\ell$= $\rm{e}$, $\mu$).
The $\mutoe$ conversion will be searched for by the DeeMe \cite{deeme2010} and COMET \cite{cometTDRNew} experiments in preparation at J-PARC. COMET will reach a sensitivity of $\mathcal{O} (10^{-15})$ in its first phase, to be compared with the extant limit of $7 \times 10^{-13}$ \cite{bertl_2006_epj} while in a second phase, it aims for a sensitivity of $\mathcal{O} (10^{-17})$. Another experiment, Mu2e \cite{Abrams:2012er} under construction at Fermilab, will search for $\mutoe$ aiming to a sensitivity of $3 \times 10^{-17}$.
The $\mutoeee$ is being searched for in a new experiment currently in the commissioning phase at PSI: Mu3e \cite{Mu3ePhaseI}, that in a staged approach plans to reach a sensitivity of $10^{-16}$, while the extant limit is $1 \times 10^{-12}$ \cite{bellgardt_1988}. 
CLFV decays involving $\uptau$ leptons will be studied at the Belle II \cite{BelleII_Phys} experiment at Super KEKB with a sensitivity goal of $\mathcal{O}(10^{-9})$.
Experiments at a super Charm-Tau factory \cite{Hao2016}, approved by the Russian government and presently in the R$\&$D phase, were proposed with a similar sensitivity reach. The sensitivity to NP in the different $\mu$ and $\uptau$ channels  depends on the specific NP models assumed, and all the above modes can be considered powerful probes to explore NP, complementary to direct searches at LHC and HL-LHC.

 The MEG II experiment, under commissioning at the PSI, aims for a sensitivity enhancement compared to the MEG final result, $\mathcal{O} (6 \times 10^{-14} )$. MEG II has the same experimental concept as MEG, which is discussed in the next section, but faces the challenge of a more intense muon beam with the need for keeping high efficiency and detector resolutions. The full MEG II proposal can be found in \cite{baldini_2018}. In this paper, we review the main aspects of the design and we report the results obtained in the engineering runs in the years 2017--2020, in view of the 2021 run that will be the first with the whole detector instrumented. An update of the expected sensitivity  is also presented.

%%%%%%%%%%%%%%%%%%%%%%%%%%%%%%%%%%%%%%%%%%
\section{Experimental Components and Methods}
\label{sec:material}
\subsection{The Experimental Approach}
In $\meg$ search experiments, positive muons are stopped in a target and the signature of a two-body decay at rest is exploited: an ${\rm e^+}$ and a $\gamma$ emitted simultaneously, moving back-to-back with their energies equal to half of the muon mass ($m_\mu/2 = 52.8$~MeV). 
The signal of the $\meg$ decay can be reconstructed by measuring the photon energy $\egamma$, the positron momentum $\ppositron$, their relative angle $\Thetaegamma$ and time difference $\tegamma$. 

The background comes from two sources: radiative muon decays (RMD, $\radiative$) with the neutrinos carrying away a small amount of energy, or from an accidental coincidence of a positron from a Michel decay ($\michel$) with a photon coming from RMD, bremsstrahlung or positron annihilation-in-flight (AIF, $\aif$). 
Since the rate of accidental background events varies quadratically with the muon stopping rate $R_{\mu^+}$, it must be chosen appropriately in order to limit the signal to background ratio and optimize the discovery potential.

\begin{figure}[t]
\includegraphics[width=10.5 cm]{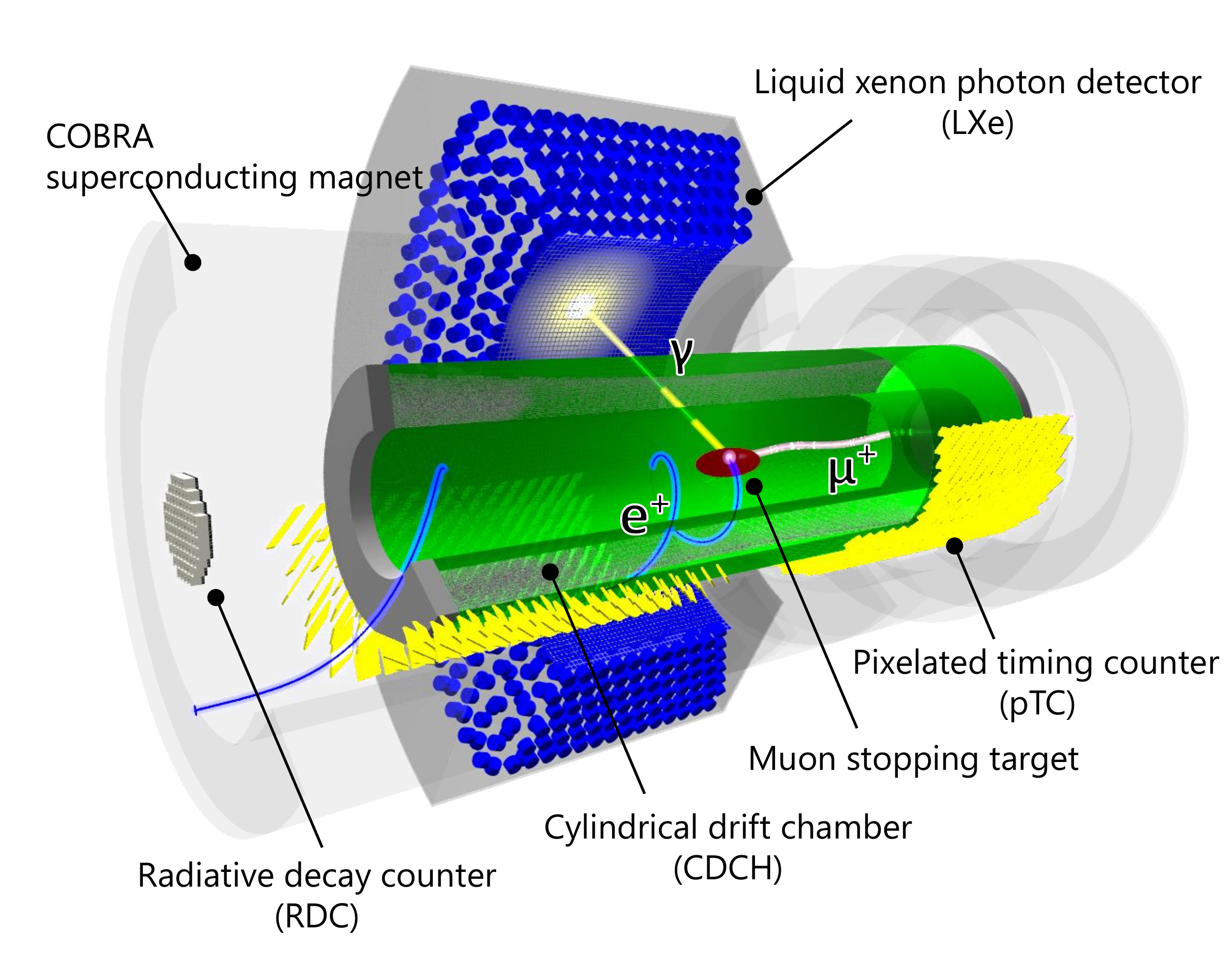}
\caption{A scheme of the MEG II experiment.
Photons and positrons from the decay of muons stopped in a target placed at the centre of the detector system are measured by the liquid xenon photon detector and the positron spectrometer consisting of a superconducting magnet, a cylindrical drift chamber, and two sets of pixelated timing counter, respectively. A new detector identifying background photons (RDC) is installed on the beam axis.}
\label{Fig2}
\end{figure}

The MEG II detector, shown schematically in Figure~\ref{Fig2}, is based on the same experimental concept as MEG and uses the same beam, but at the increased intensity of $7 \times 10^7~\mu^+$/s.

The positron spectrometer uses a gradient magnetic field 
and detects positron tracks with a low-mass single-volume cylindrical drift chamber (CDCH) with high rate capability. 
The positron time is measured by a pixelated timing counter (pTC).
The acceptance of the spectrometer is more than doubled from MEG. The photon is measured by an upgraded liquid xenon calorimeter (LXe) where the collection of scintillation light is more uniform. This is obtained by replacing the MEG photomultiplier tubes (PMTs) on the photon entrance face with smaller vacuum-ultraviolet (VUV) sensitive silicon photomultipliers (SiPMs). A novel device for active suppression of the accidental background is introduced, the radiative decay counter (RDC). 
The trigger and data-acquisition system is also upgraded to deal with an increased number of readout channels and to cope with the required bandwidth. 
Moreover, several auxiliary devices have been developed to monitor the beam and calibrate the detectors. The MEG II subsystems are described in the following subsections focusing on the changes from MEG, while the details of the MEG system can be found in \cite{megdet}.

\subsection{The Muon Beam and the Target}
Various combinations of beam momentum and target have been investigated to increase the muon stopping rate from $R_{\mu^+}=3 \times 10^7$~$\mu^+$/s   to $7 \times 10^7$~$\mu^+$/s while preserving a low material budget along the beam line and in the target region.
The configuration chosen is a surface muon beam of 28 MeV/$c$ (as in MEG) with a target of 140 $\mu$m thickness, placed at an angle of 15$^\circ$ with respect to the beam axis, which allows a simultaneous non-destructive beam intensity and profile measurement. A photogrammetric method to monitor the target position, orientation and shape has been implemented, thus addressing one of the dominant systematic effects in MEG. The system includes two CCD cameras and LEDs placed at one of the CDCH endplates, which optically monitor printed patterns (dots) on the target \cite{targetUCI}\cite{targetRoma}.

Three new detectors have been built to measure the beam profile and rate, to complement the well-established methods used in MEG, i.e., a scanning pill-counter at the collimator and a scanning APD detector at the centre of the spectrometer. The first is made by scintillating fibres and mounted at the entrance to the spectrometer. The second is a luminophore foil detector (CsI on a Mylar support) with a CCD camera installed at the intermediate focus collimator system. The third one is a MatriX detector consisting of a $9\times9$ matrix of small scintillation counters, which is used at the centre of the spectrometer.

\subsection{The Constant Bending Radius Magnet}
The COBRA (COnstant Bending RAdius) solenoid, inherited from MEG is a superconducting magnet with a graded magnetic field in the axial direction, ranging from 1.27 T at the centre to 0.49 T at both ends of the magnet cryostat. Compared to a uniform solenoidal field,  COBRA has the advantage that particles produced with small longitudinal momentum have a shorter latency time in the spectrometer, allowing operations at a high rate. Moreover, the field is designed so that positrons emitted from the target follow a trajectory with an almost constant projected bending radius, weakly dependent on the emission polar angle. The magnet is equipped with a pair of compensation coils to reduce the stray field to a suitable level for LXe PMT operations.
\subsection{The Detector}
\subsubsection{The Cylindrical Drift Chamber (CDCH)}
The new MEG II positron tracker is a cylindrical drift chamber, 191~cm long, with the radial extent ranging from 17 to 29~cm.  
%The high granularity is ensured by nine layers of 192 drift cells, each a few mm wide. 
Nine layers of 192 drift cells, each a few mm wide, allows a high granularity measurement of the positron trajectories.
The wires form an angle with the CDCH axis, varying from 6$^\circ$ in the innermost layer to 8.5$^\circ$ in the outermost one. The stereo angle depends on the layer with an alternating sign, allowing reconstruction of the longitudinal hit coordinate.
The drift cell shape is quasi-square, with a 20~$\mu$m Au-plated W sense-wire surrounded by 40 or 50~$\mu$m Ag-plated Al field wires, with a 5:1 field-to-sense wire ratio. 
The sensitive volume is filled with a low-mass He:iC$_4$H$_{10}$ (90:10) gas mixture (plus some small quantities of additives to improve the operational stability), which allows a good compromise between low scattering (radiation length $1.5 \times 10^{-3}~X_0$) and single-hit resolution (<120 $\mu$m, measured on prototypes \cite{Baldini:2016rrk}). 

Wire breaking problems arose during the CDCH assembly and commissioning, even if the operations were performed inside clean rooms under controlled environmental conditions. The problem affected aluminium wires and was investigated by performing optical inspections with microscopes, chromatography, practical tests and SEM/EDS (Scanning Electron Microscopy with Energy Dispersive Spectroscopy) analyses. The origin of the breaking phenomenon turned out to be chemical corrosion of the Al core in presence of water condensation on wires from ambient humidity. The development of corrosion was stopped by keeping the wire volume in an absolutely dry atmosphere with a continuous flow of inert gas (nitrogen or helium). 
The total number of broken cathode wires amounts to 97 (10) for diameters of 40 (50) $\mu$m.
The effect of a missing cathode wire on the reconstruction was evaluated by means of Garfield \cite{garfield} and ANSYS \cite{ansys} simulations and found to be negligible. 

The front-end electronics are constituted by custom boards, which also provide HV to the wires
and read out signals at both wire ends. The active electronics are cooled by a chiller system. The HV is supplied by a commercial HV \cite{iseg} system made of high precision modules, while a custom-made gas system \cite{Baldini:2018ing} flushes and monitors the gas mixture to the chamber and helium to COBRA.

\subsubsection{The Pixelated Timing Counter (pTC)}
The $\mathrm{e}^+$ time must be measured with a resolution of about 40 ps, at a hit rate of 5 MHz. This is needed to allow for a precise measurement of the relative time of the $\mathrm{e}^+\gamma$ pair. The $\mathrm{e}^+$ time is measured by the pTC which also generates trigger signals by providing prompt timing and direction information. 
The pTC is composed of 512 scintillator counters organized in two sectors, placed on the upstream side and downstream side of the target, each consisting of 256 counters. 
The signal $\mathrm{e}^+$, curled by the COBRA magnetic field, hits on average 9 counters, significantly improving the total time resolution compared to MEG.
The background hit rate on a single counter is less than 100~kHz even though the beam intensity is twice as high as in MEG. 

A single counter is composed of a fast plastic scintillator plate, BC-422 ($120 \times40\times5~\mathrm{mm}^ 3$ or $120 \times 50 \times 5~\mathrm{mm}^3$ depending on the counter location) and six SiPMs connected in series at each end. 
The resolution of all counters was measured with a $^{90}$Sr source and found to be below 100~ps. 
The average resolution is 72 (81)~ps for $120\times40(50)\times5~\mathrm{mm}^3$ counters. The variations of the counter resolutions come from the variations of the photon detection efficiency amongst SiPMs and variations of the light yield between the scintillators.

The pTC is equipped with a cooling system to reduce dark count rates in the SiPMs \cite{Boca:2020elp} and a laser calibration system \cite{Boca:2019bbp} to calibrate the timing between the counters.

\subsubsection{The Liquid Xenon Calorimeter (LXe)}
The liquid xenon detector allows the measurement of the photon's energy, time and position. The MEG LXe detector was one of the world's largest xenon based detectors with 900 L of liquid xenon. It was surrounded by 846 2-inch PMTs to detect the scintillation light emitted in the VUV range at a temperature of 165 K. Its performance  was limited due to the non-uniform coverage of the PMT sensitive area, especially on the photon entrance face. While we reuse the liquid xenon and the cryostat in MEG II, we have upgraded it by replacing the PMTs on the photon entrance face with smaller photosensors, in order to reduce the non-uniformity of the detector response. In addition, the layout of the PMTs on the lateral faces was rearranged to improve light collection for events near the lateral walls. 
We inherited several methods to carefully calibrate the detector from MEG, with some modifications to match the upgraded configuration. The designed energy and position resolutions are $\sim$0.5~MeV and $\sim$2.5~mm, respectively.

A new type of VUV sensitive SiPM, the VUV-MPPC \cite{IEKI2019148}, was developed. 
It has high immunity to magnetic fields and is sensitive to single photons, which enables an easier and more reliable calibration of the detector. Moreover, a finer readout granularity allows more precise reconstruction of events where the initial high energy photons convert near the entrance face (shallow events) and to reduce pile-up of these photons in the same acquisition window.
Thus, 216 PMTs at the entrance face are replaced by 4,092 SiPMs. 
A VUV-MPPC consists of four $6\times6~\mathrm{mm}^2$ sensor chips and the signals are read out by one channel per SiPM, connecting the four chips in series inside a PCB. 

\subsubsection{The Radiative Decay Counter (RDC)}
Photons contributing to accidental background come from either RMD, bremsstrahlung,  or positron AIF. The AIF background decreases in MEG II because of the reduced mass of the CDCH compared to the MEG drift chambers. The yield of the AIF background photons above 48 MeV per muon decay expected from MC simulations is 1.4 $\times$10$^{-6}$, to be compared with 2.3 $\times$ 10$^{-6}$ in MEG. On the other hand, the RMD photon background is unchanged.
The RDC, a detector newly installed in MEG II, is capable of identifying a fraction of low-energy positrons from RMD decays with photon energies close to the kinematic limit. 

The detector
consists of 12 plastic scintillators BC-418 for time measurement and 76 LYSO crystals for energy measurement. It is placed 140~cm downstream of the muon stopping target and covers the region within 10~cm from the beam axis.
It identifies low-energy positrons (1--5 MeV) in time coincidence with the detection of a high energy photon in the LXe detector. 
According to simulations, it can detect about $\sim40\%$ of RMD background with $\egamma > 48$~MeV improving the sensitivity of the $\meg$ search by $\sim10\%$. 

%Upstream RDC?
If we place another module of RDC upstream of the target, we can identify the other half of the RMD positrons emitted towards the upstream side. However, the upstream RDC is technically challenging because it must be placed in the beam path. Several detector techniques have been examined; currently, the best candidate is a resistive plate chamber with extremely low-mass diamond-like carbon based electrodes. Though the upstream RDC is not included in the baseline design of MEG II at this moment, intensive R\&D work is underway.

\subsection{The Trigger and Data Acquisition System}
The readout electronics was re-designed to deal with a factor of two increase in muon stopping rate from MEG and an almost tripling in the number of electronic channels. 
Since the physical space for the electronics could not be expanded, the only option was to combine the previous DAQ, trigger and high-voltage systems into a combined single system. 
The result is the new WaveDAQ system~\cite{Galli:2019nmv}, consisting of four distinguished subsystems.
First, the system is housed in a full custom crate with an integrated remote controlled power supply (24~V / 360~W) and a custom backplane featuring gigabit serial links in a dual star topology. 
The second system is the WaveDREAM (Drs4 based REAdout Module) board (WDB), which on one hand digitizes all input signals with a sampling frequency up to 5 GSPS / 12 bit and on the other hand performs continuous trigger operations at 80 MHz such as summing all input channels and comparing the result to a predefined threshold. In addition, the WDB can house a high voltage generator for SiPM biasing.
The third system is a dedicated Trigger Concentrator Board (TCB), which receives all digital trigger data from one crate, combines and processes them, then sends the result to a central trigger system also consisting of the same board type, where the final global trigger decision is made every 12.5 ns. 
The fourth system is the Data Concentrator Board (DCB), which receives the DAQ data stream from each WDB and sends it to the central computer via Gigabit Ethernet lines.

%%%%%%%%%%%%%%%%%%%%%%%%%%%%%%%%%%%%%%%%%
\section{Results}
\label{sec:results}
In this section the results of the engineering runs carried out in the years 2017--2020 are reported.  
%\subsecticon{The engineering runs}

During the engineering runs, the muon beam was delivered at different intensities, up to the MEG II nominal intensity ($7 \times 10^7$~$\mu^+$/s) and stopped in the target. Beam tuning and measurements were performed taking advantage of the new beam monitoring tools, which allowed continuous measurement of the beam conditions. 
The scintillating fibre and matrix detectors were frequently used during the runs, providing the beam profile and rate consistent with those measured by the pill and APD counter at the beginning of the runs. 
The target camera system has been extensively tested and showed the capability to detect target displacements at the required precision level (100~$\mu\mathrm{m}$ for movements along the direction transverse to the target plane).
In 2020, a $\uppi^-$p charge-exchange run was performed, in which the beam was changed from positive muons to 70.5 MeV/$c$ negative pions and the muon stopping target was replaced with a liquid hydrogen target to induce the reaction $\uppi^-\mathrm{p}\to\uppi^0\mathrm{n}$. Photons from the subsequent decay $\uppi^0\to \gamma \gamma$ are used for LXe calibration purposes. 

The DAQ and the trigger system operated successfully enabling stable data taking during the runs. A limited number of DAQ channels was available because the mass production
of the full electronics was not yet started, awaiting the detailed results from the
prototype test. Various  problems were identified and solved; in particular common-mode noise was reduced to a level that does not affect the photon energy resolution.  
The trigger was successfully commissioned and performed well during the runs. The full system has been deployed in spring 2021 with about 9,000 channels reading out the whole MEG II detectors. A first test showed that the system can run at an event rate of 50 Hz resulting in a data stream of about 8 GBit/s.

\begin{figure}[t]
\includegraphics[width=8. cm]{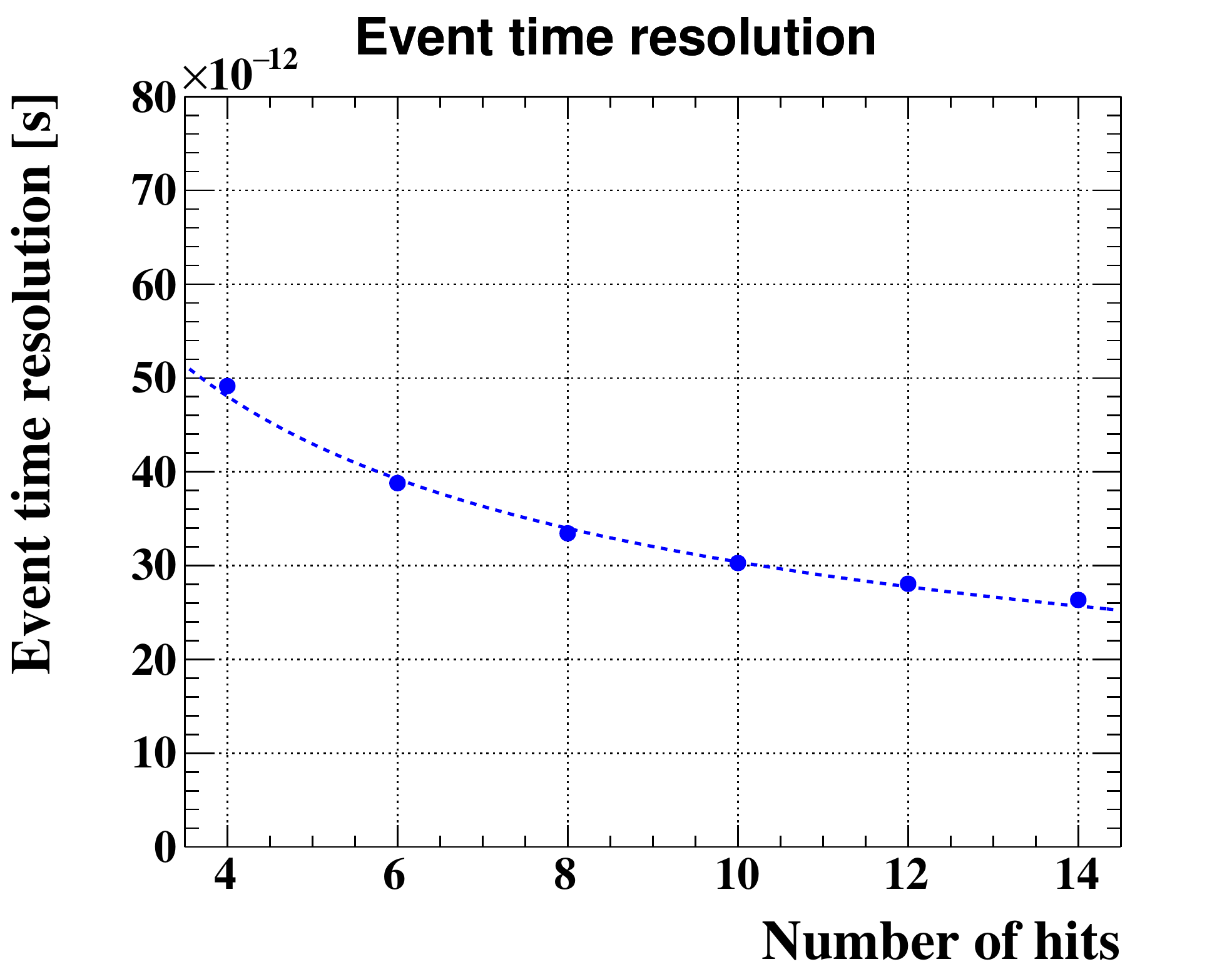}
\caption{The pTC time resolution as a function of the number of hit counters obtained from the 2017 run. 
The signal $\mathrm{e}^+$ hits nine counters on average.}
\label{FigpTCReso}
\end{figure}

The pTC was the first sub-detector that was fully commissioned already during the first engineering run, in 2017. 
It operated in the nominal MEG II muon beam. The background $\mathrm{e^+}$ hit rate was confirmed to be less than 100~kHz as expected.
Radiation damage to SiPMs was carefully studied. The dark current in the SiPMs gradually increases during the use of the beam while the impact on the time resolution is kept under control by cooling the detector to 10~$^\circ$C. 
The $\mathrm{e}^+$ time is reconstructed by combining hit times after subtracting the time-of-flight between counters. It is expected to be calculated from the $\mathrm{e}^+$ trajectory given by the CDCH but the highly segmented design of pTC allows the tracking of $\mathrm{e}^+$ using the hit pattern. This pTC-alone tracking was used to evaluate the pTC time resolution;
Figure~\ref{FigpTCReso} shows the resolution measured with the 2017 data as a function of the number of hits. 
The measured resolution is worse than the expectation from the tests with a $^{90}$Sr source made before the installation because of the larger noise level in the MEG II environment. 
By weighting with the distribution of the number of the hits for
the signal $\mathrm{e}^+$s obtained by a Monte Carlo (MC) simulation and correcting for a bias in the evaluation method, the overall time resolution is estimated to be 35~ps, a factor of two better than that in MEG (76 ps). After three years of physics run, the resolution is expected to deteriorate to 41~ps due to the radiation damage, still satisfying the requirement of about 40 ps.
In the 2018--2020 runs, only one sector, either upstream or downstream, was installed to provide a trigger to the DAQ for taking Michel e$^+$ data together with the CDCH. 

The CDCH has been integrated into the experiment since 2018. Due to the limited DAQ channels available during the engineering runs a complete particle tracking test was not possible. Nonetheless, the chamber was tested with cosmic rays and different beam intensities up to the MEG II nominal one; tests with different gas mixtures were also performed. A few wire breaks, due to corrosion that started before the chamber was put in an inert atmosphere, occurred during the runs, requiring electrical insulation of part of the chamber to continue data-taking. The broken wires had to be removed at the end of the beam periods after re-opening the detector. 
Anomalous high currents (up to 300~$\mu$A) were observed in parts of the CDCH during the runs, that were cured after optimizing the gas mixture and operating the chamber with up to 2$\%$ oxygen. 
As a result, the chamber was operated in stable conditions at the full MEG II beam intensity with He/isobutane (90/10) + isopropyl alcohol (1\%)  + O$_2$ (0.5\%) for a period of about one week during the 2020 run. Front-end (FE) electronics with three different amplifications were used to investigate the optimum electronics gain value. The total gain (gas gain $\times$ FE gain) was measured using cosmic ray data by comparing amplitude and charge spectra to MC simulations which contain a detailed description of ionization patterns, drift and diffusion from Garfield, single-electron pulse shapes from prototype measurements under UV laser light as well as the noise spectra measured on data. 
The configuration with the highest FE gain has been chosen and is currently under implementation (a modification of all the FE boards is required) since it has been demonstrated that it provides the best signal to noise ratio. The gas gain was extracted from the currents of the HV power supplies, taking into account the hit rate that was also directly measured from data. The measured gain is in the range of $4 - 7 \times 10^5$ in the mixture He/isobutane 90/10 plus 1\% isopropyl alcohol. 
Studies with simulations and data have been carried out to understand the impact of oxygen to optimize the gas mixture. 
We do not expect a significant deterioration of the chamber performances caused by the use of oxygen concentrations at the level 0.5$\%$ or below considering also the expected improvements deriving from the larger signal to noise ratio of the modified electronics.

The limited number of hits on reconstructed tracks, due to the small number of readout channels, does not allow an accurate measurement of the spatial resolution and the CDCH performances.  
Thus, the expected performances of the fully instrumented detector have been evaluated by MC simulations, updated with noise, gain and electronics response observed in data. Positron efficiency $\epsilon_{\mathrm{e}^+}$ and angular resolution $\sigma_{\theta_{\mathrm{e}^+}}$ are currently worse than the design (65$\%$ and 6.7 mrad vs.\ 70$\%$ and about 5~mrad, respectively), the vertex resolutions are already at the design level (about 1.7 mm and 0.8 mm along the beam and vertical directions, respectively) while the momentum resolution $\sigma_{p_{\mathrm{e}^+}}$ is better than the design (100~keV/$c$ vs.\ 130~keV/$c$). The efficiency loss is due to the higher than expected noise levels. We observe both low-frequency coherent noise over sets of wires on the same WaveDREAM board as well as incoherent noise extending to higher frequencies. Digital filters have been developed to cope with it, but an intensive campaign for the identification and  suppression of hardware noise sources is foreseen for the 2021 engineering run. The improved momentum resolution is due to an improved track fitting method for tracks with multiple turns in the chamber.

The upgraded LXe detector was put into operation in 2017 and calibrated with sources, LED, low-energy photons from a nuclear reaction using a Cockcroft--Walton accelerator \cite{calibration_cw} and finally with high-energy photons in the 2020 $\uppi^-$p charge-exchange run. The RMD and background spectra were acquired with the muon beam at different intensities. The long-term stability of the PMTs and MPPCs was investigated and degradation of the photosensor performance was observed under muon beam conditions. 
A PMT gain decrease at the MEG II nominal intensity was first observed to be faster than the expected based on the MEG experience in 2018, but the deterioration speed moderated year by year. 
We decided to halve the gain to further mitigate the gain decrease and the current gain decrease rate measured in 2020 (0.16$\%$/day) is now consistent with the expectation, which enables the PMTs to operate over the full MEG II data taking period.

The MPPC VUV light sensitivity (PDE) was measured to be 12$\%$ in the experiment, lower than that measured in laboratory tests, which was $>$15\%. Moreover, the PDE further decreased during the runs. In the 2020 run, the deterioration speed was lower than in 2019 (0.03~$\%$/hour instead  of 0.06~$\%$/hour, where 0.03~$\%$/hour means that the PDE will go to zero in 100/(0.03~$\%$/hour),  about 139 days.). Possible causes considered are the hole trapping at the sensor surface induced by VUV light or irradiation at low temperature. A possible solution was identified by annealing at high temperatures which removes accumulated charges. The effect was demonstrated with several MPPCs in the LXe detector; the PDEs recovered to values $>$15\%. The optimum annealing strategy for all MPPCs is currently under investigation. Since the timing resolution would be dominated by PMT timing resolution in the case of the small MPPC PDE, MC simulations showed that almost no degradation of the MEG II sensitivity occurs down to 5$\%$ PDE,
which is good enough to guarantee the sensitivity reach.
If the degradation is saturated at this value, annealing may not be necessary. 

\begin{figure}[t]
\includegraphics[width=9. cm]{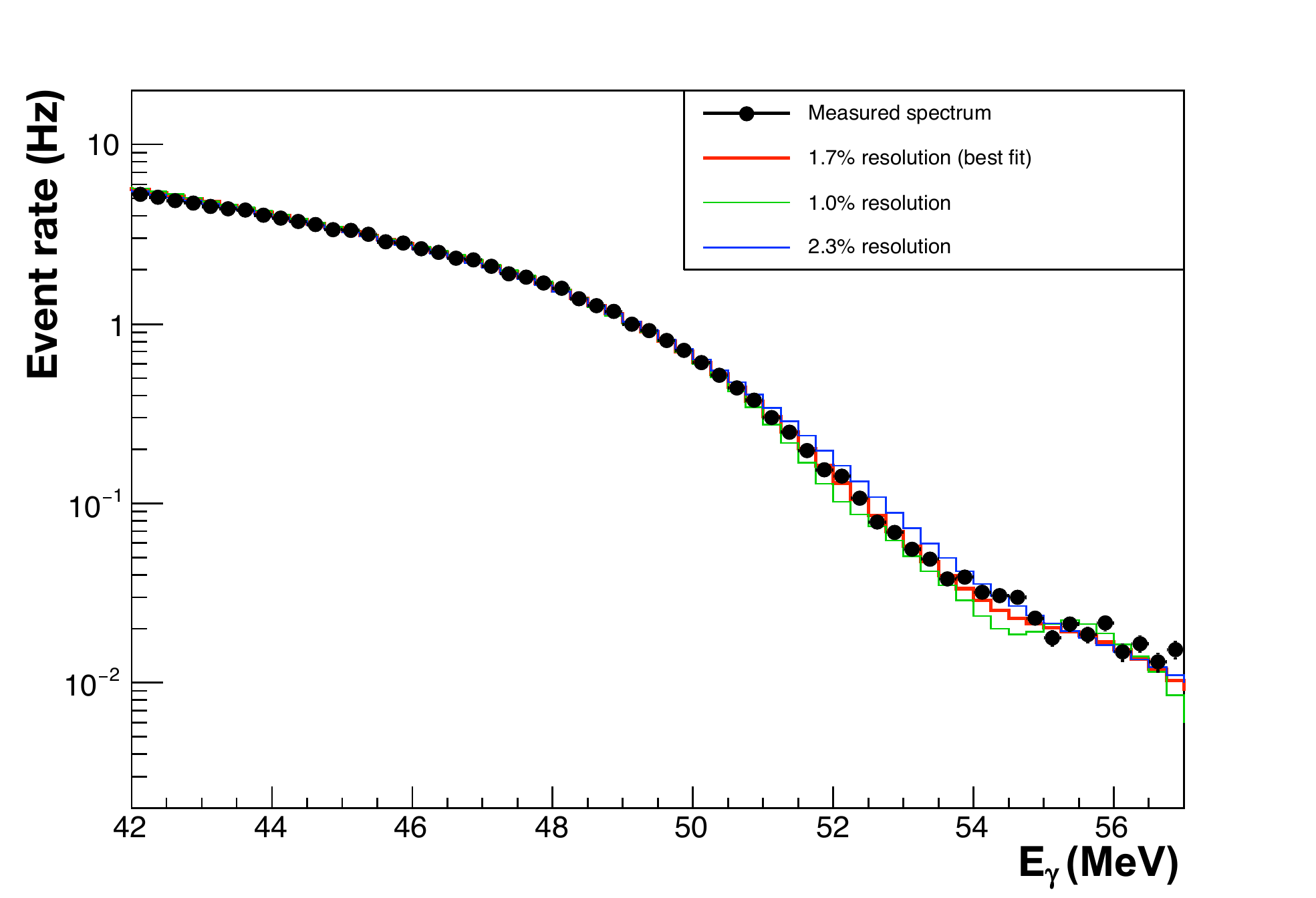}
\caption{A background photon spectrum measured in the 2019 run at the MEG II beam intensity (black points) as well as MC simulation spectra convoluted with different energy resolution assumptions. The red histogram shows the 1.7\% resolution case (best fit), and green/blue show 1.0\%/2.3\%, respectively.}
\label{lxe_energy}
\end{figure}

A preliminary estimate of the detector performance was made.
The position resolution $\sigma_{x_{\gamma}}$ was measured using photons from the $\mathrm{^7Li(p,\gamma)^8Be}$ reaction excited using the Cockcroft--Walton proton accelerator with a 5~mm wide slit collimator. Improvements in both horizontal and vertical resolution were observed ($\sigma_{x_{\gamma}}\approx 2.5$~mm), twice better than MEG. The energy resolution $\sigma_{\egamma}$ was evaluated by fitting background photon spectra from muon decays using the MC distributions as shown in Figure~\ref{lxe_energy} and from monochromatic 55~MeV photons from the $\uppi^-\mathrm{p}$ charge-exchange reaction, which provide consistent results of $\sigma_{\egamma}/\egamma= 1.7\%$ in the signal region. The improved granularity of the photon incident face almost eliminates the depth dependence of the energy resolution. The resolution seems at the level of the worst-case scenario considered in \cite{baldini_2018}; however, the data analysis is still preliminary and the results are evaluated using a limited number of electronics channels. 
The intrinsic time resolution, which was estimated by the difference of times reconstructed by even and odd channels, is 39~ps, consistent with the MC expectation. 

\begin{figure}[t]
\includegraphics[width=9. cm]{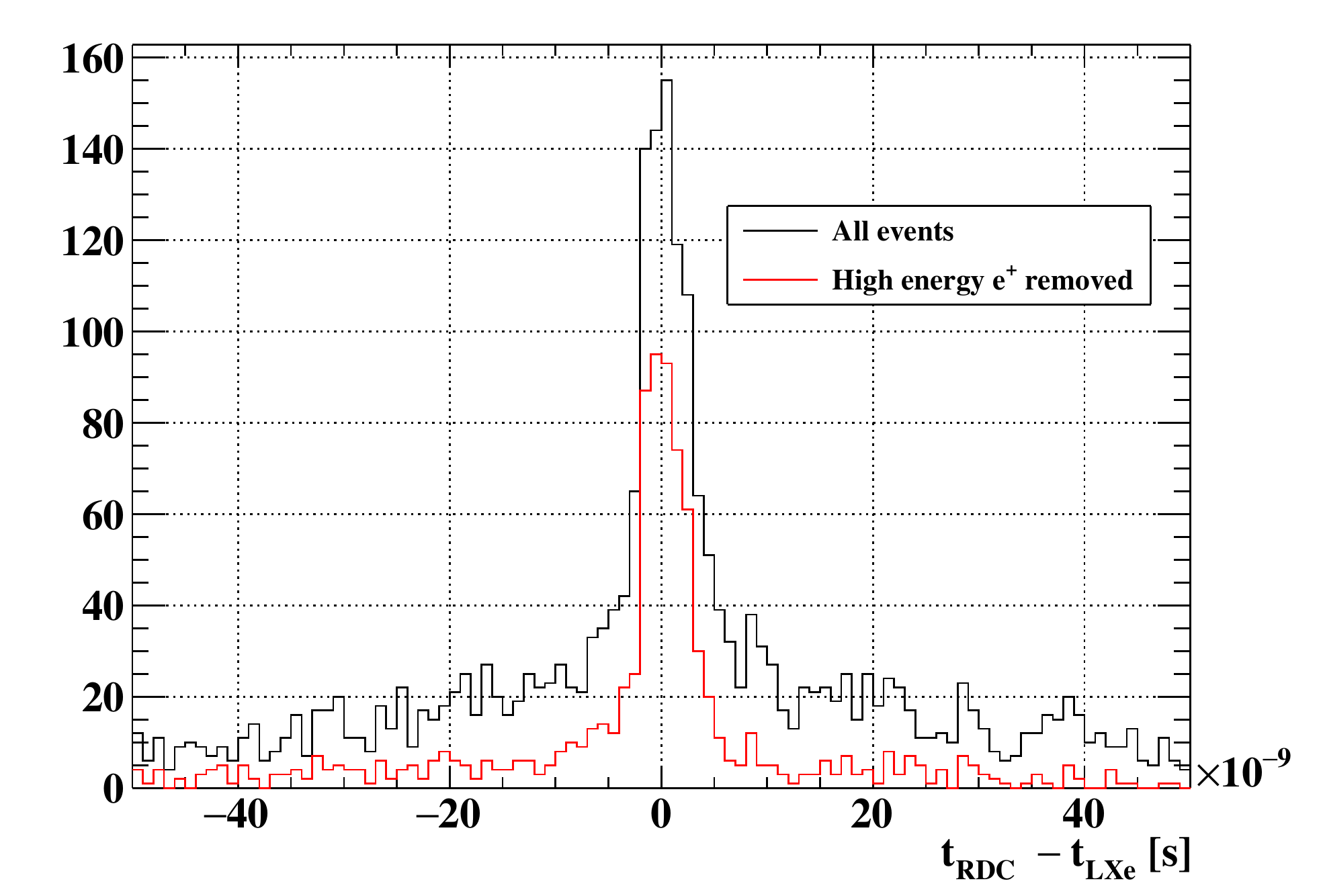}
\caption{Distribution of time difference between an $\mathrm{e}^+$ measured in RDC and a $\gamma$ in LXe detector for the events triggered by a high energy $\gamma$ at the LXe detector (black). The peak corresponds to the RMD events. The flat time distribution corresponds to accidental hits of high energy $\mathrm{e}^+$ from Michel decays, which is reduced by applying a cut on the $\mathrm{e}^+$ energy measured with the LYSO crystals of RDC ($E_\mathrm{e^+}^{\mathrm{LYSO}}<8$~MeV, red histogram).}
\label{rdc_timing}
\end{figure}

The downstream RDC 
successfully demonstrated the RMD detection capability under muon beam conditions during the engineering runs, by observing positrons in time coincidence with high energy photons in LXe. The peak in Figure~\ref{rdc_timing} represents coincident hits by a $\gamma$ and a positron from RMD. The fraction of events at the time coincident peak is found to be consistent with the expectation from simulation. An upstream RDC prototype based on resistive plate chambers with diamond-like-carbon based electrodes is being developed and showed a good efficiency and time resolution in a laboratory. The first tests under muon beam conditions were performed during the 2020 run. 

\section{Discussion}
\label{sec:discussion}

All the detector components were integrated into the MEG II experiment and tested in the engineering runs, which enables us to identify and solve issues prior to the start of physics data collection.
%\subsetction{Projected sensitivity}
Based on the results from the engineering runs, we re-evaluated the projected sensitivity of MEG II.
The updated detector performance is summarized in Table~\ref{tab:performance} (MEG II updated) 
 in comparison with that in MEG and MEG II design. We input these updated values
to the sensitivity calculation; however, note that some of the parameters were evaluated on more realistic MC simulations, updated on the basis of observation on data, and others on the data but the results are still preliminary.
The following is assumed here:
\begin{itemize}
	\item The DAQ time is 20 weeks per year with 84\% live time for three years;
	\item The muon stopping rate is $R_{\mu^+}=7\times 10^7$~$\mu^+$/s;
	\item The PDE of VUV-MPPCs is at a constant value of 6\%;
	\item The upstream RDC is not included.
\end{itemize}
The \emph{sensitivity} ($S_{90}$) is defined as the median of 90\% confidence branching-ratio upper limits
for the background-only hypothesis.
In three years, we will reach a sensitivity of $S_{90} = 6\times10^{-14}$, which is roughly one order of magnitude better than MEG's sensitivity of $5.3\times10^{-13}$. The expected single event sensitivity is $1\times10^{-14}$. In the realistic evaluation of the detector performance, some parameters improved while others became worse than the assumed design parameters; however, the overall projected sensitivity stays almost the same as that reported in Ref.~\cite{baldini_2018}. The data statistics is 5.6 times larger than MEG, and the remaining improvement of a factor of 1.6 comes from the lower background due to the improved detector resolutions. Actually, the sensitivity improvement is not yet saturated and an extension of DAQ time can improve the sensitivity reach further. 

If the degradation of VUV-MPPC PDE is not saturated at a value above 5\%, we have to recover it by the annealing process. Because annealing all the MPPCs takes a couple of months, it can be done only during the annual accelerator shutdown period. In this case, the obtainable statistics is limited by the accumulated beam intensity up to which the MPPC is operational. For the given statistics, it is better to run for the full beam time at a reduced beam intensity since the number of accidental background events has a quadratic dependence on the beam intensity. Moreover, the detector performance is better with a lower beam intensity due to a low pileup probability. For example, the positron efficiency improves from 65\% at $7\times 10^7$~$\mu^+$/s to 74\% at $3.5\times 10^7$~$\mu^+$/s with the current reconstruction algorithms. 
As a result, it was shown by a preliminary study that a comparable sensitivity is achievable at a stopping rate down to $3.5\times 10^7$~$\mu^+$/s, with which the VUV-MPPCs can operate for 20 weeks of DAQ based on the degradation speed observed in the engineering runs.

% The MDPI table float is called specialtable
\end{paracol}
\nointerlineskip
\begin{specialtable}[tb] 
\widetable
\caption{Summary of detector performance and sensitivity. 
$\sigma_X$ denotes the resolution of variable $X$ and $\epsilon_Y$ denotes the detection efficiency of particle $Y$.
Two values for $\sigma_{\egamma}$ are values for shallow ($<$2~cm) / deep ($>$2~cm) events. $\sigma_{\tegamma}$ is the coincidence time resolution for an $\mathrm{e^+}\gamma$ pair and evaluated by combining the resolutions of the positron spectrometer and the LXe photon detector. $S_{90}$ denotes the sensitivity of the experiment; see the text for the definition.\label{tab:performance}}
\tablesize{\small} %% You can specify the fontsize here, e.g., \tablesize{\footnotesize}. If commented out \small will be used.
\begin{tabular}{lccccccccc}
\toprule
\textbf{}	
& \mbox{\boldmath $R_{\mu^+}$}
& \mbox{\boldmath $\sigma_{\ppositron}$}	
& \mbox{\boldmath $\sigma_{\theta_\mathrm{e^+}}$}
& \mbox{\boldmath $\sigma_{\egamma}$}
& \mbox{\boldmath $\sigma_{x_\gamma}$} 
& \mbox{\boldmath $\sigma_{\tegamma}$} 
& \mbox{\boldmath $\epsilon_{\mathrm{e^+}}$} 
& \mbox{\boldmath $\epsilon_{\gamma}$} 
& \mbox{\boldmath $S_{90}$}	\\
\midrule
MEG & $3\times 10^7~\mathrm{s^{-1}}$ & 380~keV/$c$	& 9.4~mrad & 2.4\%/1.7\% & 5~mm & 122~ps & 30\% & 63\% & $5.3 \times 10^{-13}$\\
MEG II design & $7\times 10^7~\mathrm{s^{-1}}$ & 130~keV/$c$	& 5.3~mrad & 1.1\%/1.0\% & 2.4~mm & 84~ps & 70\% & 69\% & $6 \times 10^{-14}$\\
MEG II updated & $7\times 10^7~\mathrm{s^{-1}}$ & 100~keV/$c$	& 6.7~mrad & 1.7\%/1.7\% & 2.4~mm & 70~ps & 65\% & 69\% & $6 \times 10^{-14}$\\
\bottomrule
\end{tabular}
\end{specialtable}
\begin{paracol}{2}
\linenumbers
\switchcolumn

% new CDCH
Cathode and field wires of the CDCH suffer from corrosion caused by exposure to humidity during construction. The risk of possible wire breakings during data taking imposed the construction of a new backup chamber with different cathode and field wires. The new chamber will use more robust 50 $\mu$m diameter Ag-coated aluminium wires, instead of the old 40 $\mu$m ones,  for which the final drawing process, which was verified to weaken the wires and make them more prone to corrosion, will be avoided.
Construction of the new chamber will take approximately 18 months; in the meanwhile, the existing one will be used in the experiment.

% prospects
At the time of writing, preparation for the 2021 engineering run is underway. In this run, all the detector signals will be collected for the first time with the full readout electronics. The stability and performance of the existing CDCH and the long-term behaviour of the PDE degradation of the VUV-MPPCs will be extensively studied. We plan to collect data with a $\meg$ trigger at the end of the run, which will be the first physics data of MEG II.

%%%%%%%%%%%%%%%%%%%%%%%%%%%%%%%%%%%%%%%%%%
%\section{Patents}

%This section is not mandatory, but may be added if there are patents resulting from the work reported in this manuscript.

%%%%%%%%%%%%%%%%%%%%%%%%%%%%%%%%%%%%%%%%%%
\vspace{6pt}

\funding{
This research was funded by  
DOE DEFG02-91ER40679 (USA), 
INFN, MIUR Montalcini D.M. 2014 n.~975 (Italy),
JSPS Core-to-Core Program, A. Advanced Research Networks JPJSCCA20180004, 
JSPS KAKENHI Grant Numbers JP26000004, JP18K13557, JP19J13635, JP19J21535, JP19J21730, JP20H00154, JP21H00065 (Japan),
the Russian Federation Ministry of Science and Higher Education (Russia), 
and Schweizerischer Nationalfonds (SNF) Grant 200020$\_$172706 and Grant 206021$\_$177038 (Switzerland).
}

\acknowledgments{We are grateful for the support and co-operation provided 
by PSI as the host laboratory and to the technical and 
engineering staff of our institutes.}

\conflictsofinterest{The authors declare no conflict of interest. %Authors must identify and declare any personal circumstances or interest that may be perceived as inappropriately influencing the representation or interpretation of reported research results. Any role of the funders in the design of the study; in the collection, analyses or interpretation of data; in the writing of the manuscript, or in the decision to publish the results must be declared in this section. If there is no role, please state ``The funders had no role in the design of the study; in the collection, analyses, or interpretation of data; in the writing of the manuscript, or in the decision to publish the~results''.
} 

\end{paracol}

\reftitle{References}


\externalbibliography{yes}
\bibliography{MEG}

\begin{thebibliography}{999}

\bibitem[Calibbi and Signorelli(2018)]{Cali2018}
Calibbi, L.; Signorelli, G.
\newblock Charged Lepton Flavor Violation: an experimental and theoretical
  introduction.
\newblock {\em Riv. Nuovo Cimento} {\bf 2018}, {\em 41},~71--174,
  \href{http://arxiv.org/abs/1709.00294}{{\normalfont
  [arXiv:hep-ph/1709.00294]}}.
\newblock
  doi:{\changeurlcolor{black}\href{https://doi.org/10.1393/ncr/i2018-10144-0}{\detokenize{10.1393/ncr/i2018-10144-0}}}.

\bibitem[Abi \em{et~al.}(2021)Abi, Albahri, Al-Kilani, Allspach, Alonzi,
  Anastasi, et~al.]{gminus2:2021}
Abi, B.; Albahri, T.; Al-Kilani, S. et~al.
\newblock Measurement of the Positive Muon Anomalous Magnetic Moment to 0.46
  ppm.
\newblock {\em Phys. Rev. Lett.} {\bf 2021}, {\em 126},~141801.
\newblock
  doi:{\changeurlcolor{black}\href{https://doi.org/10.1103/PhysRevLett.126.141801}{\detokenize{10.1103/PhysRevLett.126.141801}}}.

\bibitem[Calibbi \em{et~al.}(2021)Calibbi, L\'{o}pez-Ib\'{a}\~{n}ez, Melis, and
  Vives]{Calibbi2021}
Calibbi, L.; L\'{o}pez-Ib\'{a}\~{n}ez, M.L.; Melis, A. et~al.
\newblock {Implications of the Muon g-2 result on the flavour structure of the
  lepton mass matrix},  2021,
  \href{http://arxiv.org/abs/2104.03296}{{\normalfont
  [arXiv:hep-ph/2104.03296]}}.

\bibitem[Lindner and Queiroz(2018)]{Lindner2018}
Lindner, M.;~Platscher, M.; Queiroz, F.
\newblock {A call for new physics: The muon anomalous magnetic moment and
  lepton flavor violation}.
\newblock {\em Phys. Rep.} {\bf 2018}, {\em 731},~1--82,
  \href{http://arxiv.org/abs/1610.06587}{{\normalfont
  [arXiv:hep-ph/1610.06587]}}.

\bibitem[Baldini \em{et~al.}(2016)Baldini, Bao, Baracchini, et~al.]{meg2016}
Baldini, A.M.; Bao, Y.; Baracchini, E. et~al. (MEG Collaboration).
\newblock {Search for the lepton flavour violating decay $\meg$ with the full
  dataset of the MEG experiment}.
\newblock {\em Eur. Phys. J. C} {\bf 2016}, {\em 76},~434,
  \href{http://arxiv.org/abs/1605.05081}{{\normalfont
  [arXiv:hep-ex/1605.05081]}}.
\newblock
  doi:{\changeurlcolor{black}\href{https://doi.org/10.1140/epjc/s10052-016-4271-x}{\detokenize{10.1140/epjc/s10052-016-4271-x}}}.

\bibitem[Kinsho \em{et~al.}(2010)Kinsho, Ikegami, Kawamura, et~al.]{deeme2010}
Kinsho, M.; Ikegami, M.; Kawamura, N. et~al.
\newblock Proposal of an Experimental Search for \convme\ Conversion in Nuclear
  Field at Sensitivity of $10^{-14}$ with Pulsed Proton Beam from {RCS}.
\newblock
  https://j-parc.jp/researcher/Hadron/en/pac\_1101/pdf/KEK\_J-PARC-PAC2010-13.pdf,
   2010.

\bibitem[Abramishvili \em{et~al.}(2020)Abramishvili, Adamov, Akhmetshin, Allin,
  Angélique, Anishchik, Aoki, et~al.]{cometTDRNew}
Abramishvili, R.; Adamov, G.; Akhmetshin, R.R. et~al. ({COMET} Collaboration).
\newblock {COMET} Phase-{I} Technical Design Report.
\newblock {\em Prog. Theor. Exp. Phys.} {\bf 2020}, {\em 2020},~033C01.
\newblock
  doi:{\changeurlcolor{black}\href{https://doi.org/10.1093/ptep/ptz125}{\detokenize{10.1093/ptep/ptz125}}}.

\bibitem[Bertl \em{et~al.}(2006)Bertl, Engfer, Hermes, Kurz, Kozlowski, Kuth,
  Otter, Rosenbaum, Ryskulov, van~der Schaaf, Wintz, Zychor, and {The SINDRUM
  II Collaboration}]{bertl_2006_epj}
Bertl, W.; Engfer, R.; Hermes, E. et~al. (SINDRUM II Collaboration).
\newblock A search for \convme\ conversion in muonic gold.
\newblock {\em Eur. Phys. J. C} {\bf 2006}, {\em 47},~337--346.
\newblock
  doi:{\changeurlcolor{black}\href{https://doi.org/10.1140/epjc/s2006-02582-x}{\detokenize{10.1140/epjc/s2006-02582-x}}}.

\bibitem[Abrams \em{et~al.}(2012)Abrams, Alezander, Ambrosio,
  et~al.]{Abrams:2012er}
Abrams, R.; Alezander, D.; Ambrosio, G. et~al. (Mu2e Collaboration).
\newblock {Mu2e Conceptual Design Report},  2012,
  \href{http://arxiv.org/abs/1211.7019}{{\normalfont
  [arXiv:physics.ins-det/1211.7019]}}.

\bibitem[Arndt \em{et~al.}(2021)Arndt, Augustin, Baesso, et~al.]{Mu3ePhaseI}
Arndt, K.; Augustin, H.; Baesso, P. et~al.
\newblock Technical design of the phase I Mu3e experiment,  2021,
  \href{http://arxiv.org/abs/2009.11690}{{\normalfont
  [arXiv:physics.ins-det]/2009.11690]}}.

\bibitem[Bellgardt \em{et~al.}(1988)Bellgardt, Otter, Eichler, Felawka,
  Niebuhr, Walter, Bertl, Lordong, Martino, Egli, Engfer, Grab,
  Grossmann-Handschin, Hermes, Kraus, Muheim, Pruys, Schaaf, and
  Vermeulen]{bellgardt_1988}
Bellgardt, U.; Otter, G.; Eichler, R. et~al.
\newblock Search for the decay \mute.
\newblock {\em Nucl. Phys. B} {\bf 1988}, {\em 299},~1--6.
\newblock
  doi:{\changeurlcolor{black}\href{https://doi.org/10.1016/0550-3213(88)90462-2}{\detokenize{10.1016/0550-3213(88)90462-2}}}.

\bibitem[Kou \em{et~al.}(2019)Kou, Urquijo, Altmannshofer,
  et~al.]{BelleII_Phys}
Kou, E.; Urquijo, P.; Altmannshofer, W. et~al.
\newblock The {Belle II} Physics Book.
\newblock {\em Prog. Theor. Exp. Phys.} {\bf 2019}, {\em 2019},~123C01.
\newblock
  doi:{\changeurlcolor{black}\href{https://doi.org/10.1093/ptep/ptz106}{\detokenize{10.1093/ptep/ptz106}}}.

\bibitem[Hao \em{et~al.}(2016)Hao, Ren-You, Liang, Wen-Gan, Lei, and
  Chong]{Hao2016}
Hao, Z.; Ren-You, Z.; Liang, H. et~al.
\newblock Searching for $\tmg$ lepton-flavor-violating decay at super Charm-Tau
  factory.
\newblock {\em Eur. Phys. J. C} {\bf 2016}, {\em 76},~421,
  \href{http://arxiv.org/abs/1602.01181}{{\normalfont
  [arXiv:hep-ph/1602.01181]}}.
\newblock
  doi:{\changeurlcolor{black}\href{https://doi.org/10.1140/epjc/s10052-016-4251-1}{\detokenize{10.1140/epjc/s10052-016-4251-1}}}.

\bibitem[Baldini \em{et~al.}(2018)Baldini, Baracchini, Bemporad, Berg,
  Biasotti, Boca, Cattaneo, Cavoto, Cei, Chiappini, Chiarello, Chiri, Cocciolo,
  Corvaglia, de~Bari, {De Gerone}, D'Onofrio, Francesconi, Fujii, Galli, Gatti,
  Grancagnolo, Grassi, Grigoriev, Hildebrandt, Hodge, Ieki, Ignatov, Iwai,
  Iwamoto, Kaneko, Kasami, Kettle, Khazin, Khomutov, Korenchenko, Kravchuk,
  Libeiro, Maki, Matsuzawa, Mihara, Milgie, Molzon, Mori, Morsani,
  Mtchedilishvili, Nakao, Nakaura, Nicol{\`{o}}, Nishiguchi, Nishimura, Ogawa,
  Ootani, Panareo, Papa, Pepino, Piredda, Popov, Raffaelli, Renga, Ripiccini,
  Ritt, Rossella, Rutar, Sawada, Signorelli, Simonetta, Tassielli, Uchiyama,
  Usami, Venturini, Voena, Yoshida, Yudin, and Zhang]{baldini_2018}
Baldini, A.M.; Baracchini, E.; Bemporad, C. et~al. (MEG II Collaboration).
\newblock {The design of the MEG II experiment}.
\newblock {\em Eur. Phys. J. C} {\bf 2018}, {\em 78},~380.
\newblock
  doi:{\changeurlcolor{black}\href{https://doi.org/10.1140/epjc/s10052-018-5845-6}{\detokenize{10.1140/epjc/s10052-018-5845-6}}}.

\bibitem[Adam \em{et~al.}(2013)Adam, Bai, Baldini, Baracchini, et~al.]{megdet}
Adam, J.; Bai, X.; Baldini, A.M. et~al.
\newblock {The MEG detector for \megc\ decay search}.
\newblock {\em Eur. Phys. J. C} {\bf 2013}, {\em 73},~2365,
  \href{http://arxiv.org/abs/1303.2348}{{\normalfont
  [arXiv:physics.ins-det/1303.2348]}}.
\newblock
  doi:{\changeurlcolor{black}\href{https://doi.org/10.1140/epjc/s10052-013-2365-2}{\detokenize{10.1140/epjc/s10052-013-2365-2}}}.

\bibitem[Palo \em{et~al.}(2019)Palo, Hildebrandt, Hofer, Kyle,
  et~al.]{targetUCI}
Palo, D.; Hildebrandt, M.; Hofer, A. et~al.
\newblock Precise Photographic Monitoring of MEG II Thin-film Muon Stopping
  Target Position and Shape.
\newblock {\em Nucl. Instrum. Methods A} {\bf 2019}, {\em 944},~162511.
\newblock
  doi:{\changeurlcolor{black}\href{https://doi.org/10.1016/j.nima.2019.162511}{\detokenize{10.1016/j.nima.2019.162511}}}.

\bibitem[Cavoto \em{et~al.}(2021)Cavoto, Chiarello, Hildebrandt, Hofer,
  et~al.]{targetRoma}
Cavoto, G.; Chiarello, G.; Hildebrandt, M. et~al.
\newblock A photogrammetric method for target monitoring inside the MEG II
  detector.
\newblock {\em Rev. Sci. Instrum.} {\bf 2021}, {\em 92},~043707.
\newblock
  doi:{\changeurlcolor{black}\href{https://doi.org/10.1063/5.0034842}{\detokenize{10.1063/5.0034842}}}.

\bibitem[Baldini \em{et~al.}(2016)Baldini et~al.]{Baldini:2016rrk}
Baldini, A.M. et~al.
\newblock {Single-hit resolution measurement with MEG II drift chamber
  prototypes}.
\newblock {\em J. Instrum.} {\bf 2016}, {\em 11},~P07011,
  \href{http://arxiv.org/abs/1605.07970}{{\normalfont
  [arXiv:physics.ins-det/1605.07970]}}.
\newblock
  doi:{\changeurlcolor{black}\href{https://doi.org/10.1088/1748-0221/11/07/P07011}{\detokenize{10.1088/1748-0221/11/07/P07011}}}.

\bibitem[gar()]{garfield}
{Garfield++ -- simulation of gaseous detectors}.
\newblock \url{https://garfieldpp.web.cern.ch/garfieldpp/}.

\bibitem[ans()]{ansys}
{\em Ansys® Electronic Desktop 2019.}

\bibitem[ise()]{iseg}
{ISEG EHS 8630p-305F}.
\newblock \url{http://www.iseg-hv.com}.

\bibitem[Baldini \em{et~al.}(2018)Baldini, Baracchini, Cavoto, Cei,
  et~al.]{Baldini:2018ing}
Baldini, A.M.; Baracchini, E.; Cavoto, G. et~al.
\newblock {Gas Distribution and Monitoring for the Drift Chamber of the MEG-II
  Experiment}.
\newblock {\em J. Instrum.} {\bf 2018}, {\em 13},~P06018,
  \href{http://arxiv.org/abs/1804.08482}{{\normalfont
  [arXiv:physics.ins-det/1804.08482]}}.
\newblock
  doi:{\changeurlcolor{black}\href{https://doi.org/10.1088/1748-0221/13/06/P06018}{\detokenize{10.1088/1748-0221/13/06/P06018}}}.

\bibitem[Boca \em{et~al.}(2021)Boca, Cattaneo, De~Gerone, et~al.]{Boca:2020elp}
Boca, G.; Cattaneo, P.W.; De~Gerone, M. et~al.
\newblock {Timing resolution of a plastic scintillator counter read out by
  radiation damaged SiPMs connected in series}.
\newblock {\em Nucl. Instrum. Methods A} {\bf 2021}, {\em 999},~165173,
  \href{http://arxiv.org/abs/2005.05027}{{\normalfont
  [arXiv:physics.ins-det/2005.05027]}}.
\newblock
  doi:{\changeurlcolor{black}\href{https://doi.org/10.1016/j.nima.2021.165173}{\detokenize{10.1016/j.nima.2021.165173}}}.

\bibitem[Boca \em{et~al.}(2019)Boca, Cattaneo, De~Gerone, et~al.]{Boca:2019bbp}
Boca, G.; Cattaneo, P.W.; De~Gerone, M. et~al.
\newblock {The laser-based time calibration system for the MEG II pixelated
  Timing Counter}.
\newblock {\em Nucl. Instrum. Methods A} {\bf 2019}, {\em 947},~162672,
  \href{http://arxiv.org/abs/1907.00911}{{\normalfont
  [arXiv:physics.ins-det/1907.00911]}}.
\newblock
  doi:{\changeurlcolor{black}\href{https://doi.org/10.1016/j.nima.2019.162672}{\detokenize{10.1016/j.nima.2019.162672}}}.

\bibitem[Ieki \em{et~al.}(2019)Ieki, Iwamoto, Kaneko, Kobayashi, Matsuzawa,
  Mori, Ogawa, Onda, Ootani, Sawada, Sato, and Yamada]{IEKI2019148}
Ieki, K.; Iwamoto, T.; Kaneko, D. et~al.
\newblock Large-area MPPC with enhanced VUV sensitivity for liquid xenon
  scintillation detector.
\newblock {\em Nucl. Instrum. Methods A} {\bf 2019}, {\em 925},~148--155.
\newblock
  doi:{\changeurlcolor{black}\href{https://doi.org/10.1016/j.nima.2019.02.010}{\detokenize{10.1016/j.nima.2019.02.010}}}.

\bibitem[Galli \em{et~al.}(2019)Galli, Baldini, Cei, Chiappini,
  et~al.]{Galli:2019nmv}
Galli, L.; Baldini, A.M.; Cei, F. et~al.
\newblock {WaveDAQ}: An highly integrated trigger and data acquisition system.
\newblock {\em Nucl. Instrum. Methods A} {\bf 2019}, {\em 936},~399--400.
\newblock
  doi:{\changeurlcolor{black}\href{https://doi.org/10.1016/j.nima.2018.07.067}{\detokenize{10.1016/j.nima.2018.07.067}}}.

\bibitem[Adam \em{et~al.}(2011)Adam, Bai, Baldini, Baracchini,
  et~al.]{calibration_cw}
Adam, J.; Bai, X.; Baldini, A. et~al. (MEG Collaboration).
\newblock {Calibration and monitoring of the {MEG} experiment by a proton beam
  from a {C}ockcroft--{W}alton accelerator}.
\newblock {\em Nucl. Instrum. Methods A} {\bf 2011}, {\em 641},~19--32.
\newblock
  doi:{\changeurlcolor{black}\href{https://doi.org/10.1016/j.nima.2011.03.048}{\detokenize{10.1016/j.nima.2011.03.048}}}.

\end{thebibliography}

% If authors have biography, please use the format below
%\section*{Short Biography of Authors}
%\bio
%{\raisebox{-0.35cm}{\includegraphics[width=3.5cm,height=5.3cm,clip,keepaspectratio]{Definitions/author1.pdf}}}
%{\textbf{Firstname Lastname} Biography of first author}
%
%\bio
%{\raisebox{-0.35cm}{\includegraphics[width=3.5cm,height=5.3cm,clip,keepaspectratio]{Definitions/author2.jpg}}}
%{\textbf{Firstname Lastname} Biography of second author}

% The following MDPI journals use author-date citation: Arts, Econometrics, Economies, Genealogy, Humanities, IJFS, JRFM, Laws, Religions, Risks, Social Sciences. For those journals, please follow the formatting guidelines on http://www.mdpi.com/authors/references
% To cite two works by the same author: \citeauthor{ref-journal-1a} (\citeyear{ref-journal-1a}, \citeyear{ref-journal-1b}). This produces: Whittaker (1967, 1975)
% To cite two works by the same author with specific pages: \citeauthor{ref-journal-3a} (\citeyear{ref-journal-3a}, p. 328; \citeyear{ref-journal-3b}, p.475). This produces: Wong (1999, p. 328; 2000, p. 475)

%%%%%%%%%%%%%%%%%%%%%%%%%%%%%%%%%%%%%%%%%%
%% for journal Sci
%\reviewreports{\\
%Reviewer 1 comments and authors’ response\\
%Reviewer 2 comments and authors’ response\\
%Reviewer 3 comments and authors’ response
%}
%%%%%%%%%%%%%%%%%%%%%%%%%%%%%%%%%%%%%%%%%%
\end{document}